# A Generalized, Frequency Domain-Embedded, State Space Approach for Controller Synthesis

**Ahmad A. Masoud**

Electrical Engineering Department, The Center for Communication Systems and Sensing, KFUPM, Daharan, Saudi Arabia, e-mail: masoud@kfupm.edu.sa

**ABSTRACT**  This paper develops a novel control synthesis approach for a wide class of practical systems. The control action is derived by inserting a compensator device in the forward path of the system that is to be controlled. The compensator design method is based on a state space system that is embedded in the frequency domain of the plant. It uses a nonlinear subspace that is the image of the compensator being used to determine the parameters of the compensator and modify system behavior. The coordinates of the state space system are taken as the compensator's parameters. The method is capable of designing a compensator of arbitrary order to make the system comply with given stability/performance requirements provided that these requirements can be geometrically-interpreted in the frequency domain. The approach is developed and a proof of its ability to converge, if a solution exists, to the compensator's tuning parameter set that satisfy the desired performance conditions is provided. A set of design examples are supplied to demonstrate the applicability of the approach to different types of linear, nonlinear, SISO and MIMO systems and system with delays.

## I. INTRODUCTION

Frequency domain methods [1,2,3] are powerful and practical tools for analyzing the performance and stability of a wide class of systems. Their ability to provide, in the frequency domain (FD), a geometric interpretation of the stability and performance conditions has many advantages. Among other things, it makes it possible to:

1- Isolate system analysis from its order. In frequency domain methods, a first order system is analyzed in the same manner as an N-order system (N>1). An important consequence of this feature is the applicability of frequency domain methods to any type of dynamical systems that may be represented as a rational function (e.g. fractional order systems and systems with delays, to mention a few),

2- Frequency domain methods can analyze, in a unified manner, systems that are of different nature. Conclusions about system behavior may be derived from the state of its frequency response relative to the geometric region, which encodes the stability/performance constraints. It is well reported in the literature that FD methods can be applied to linear systems [4,5], nonlinear systems [6,7,8,9], systems with delays [10,11] and multi-input multi-output (MIMO) systems [12],

3- The manner in which system behavior is inferred is invariant to its structure or the location in that system of the component whose effect on behavior is to be examined. This feature enables the utilization of frequency domain analysis techniques in the construction of flexible and versatile control synthesis methods that can handle a wide class of challenging dynamical systems.

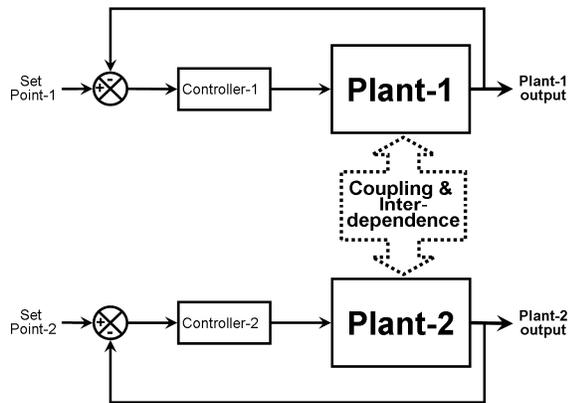

Figure-1: Compensators inserted in cascade with plants to be controlled

In frequency domain methods, system behavior is inferred based on its frequency response relative to the geometric region encoding stability/performance constraints. It does not matter whether the spectrum modifiers are linear or nonlinear as long as the frequency response of the system avoids the undesired regions. Therefore, the control synthesis approach adopted in this paper is based on adjusting the spectrum of the system using a linear system (compensator) that is cascaded in series with the plant. In addition, the suggested approach places no restriction on the order of the compensator that maybe tackled. This provides the compensator with high capabilities to adjust the spectrum of the system and satisfy the geometric constraints that encode the stability and performance requirements even if they act on an interconnected MIMO system in a decentralized manner (Figure-1).

Many methods for compensator-based, spectrum shaping were suggested [4,5]. Although these methods have proven effective in synthesizing practical controllers, they do suffer shortcomings that limit their utility and the types of systems they may be applied to. For example, they can mainly design low order compensators (1st or 2nd order) or at best cascade 1st order compensators to create a higher order one. Cascading of low order compensators imposes stringent constraints on the tuning parameters of the overall compensator. This could lead to the exclusion of a useful set of behaviors from those which the compensator has the potential of inducing. These synthesis techniques seem to be mostly based on the (-1,0)-centered circle criterion or aspects of it (e.g. phase or gain margins). Some frequency domain stability criteria, such as the Popov criterion [13,14,15,16], are not based on circular regions. It is important that a frequency domain synthesis method produce a compensator that is able to make the frequency response comply with an arbitrary region in the FD. It is also advantageous to explicitly factor-in the value of the frequency in the frequency response shaping process. This makes it possible to derive realizable frequency domain criteria that cover a wide class of situations. While spectrum shaping (loop shaping) techniques exist for the MIMO case [17,18], they don't seem to avail themselves of an explicit frequency domain geometry. Most of the geometry-explicit spectrum shaping techniques are focused on single-input single-output (SISO) systems.

The suggested approach avoids the limitations of existing compensator design techniques and achieves the flexibility needed to accommodate arbitrary geometry of the stability/performance regions which the frequency response needs to avoid. It does so by hybridizing the frequency domain approach with the state space approach. The frequency response of the system is affected by virtual actions that jointly attempt to satisfy the geometric constraints and lie in the subspace of actions that the compensator can exert (i.e. the action is realizable by the compensator).

State space control techniques [19] are commonly viewed as alternatives to frequency domain methods. However, due to the flexibility in defining what a state is and the abstract nature of a space, it is possible to hybridize the two techniques together. The hybrid is constructed so that the state space approach operates within the confines of the frequency domain approach. The notion of a state is based on whether or not the frequency response of the system complies with the geometric stability/performance criterion. Compliance is quantified by a scalar function covering frequency domain. The function measures the amount of intersection the frequency response makes with the forbidden FD regions, which causes the system to violate the stability conditions. The components of the state vector are values of this function taken at samples of the system's frequency response. These samples may be selected so that it is possible to judge from the samples of the state vector the condition of the whole frequency response. The elements of the state space system input vector are the parameters used to tune the compensator. The design of the compensator that would make the frequency response comply with stability/performance conditions reduces to finding a stabilizing controller for the state of the embedded state space system. This controller is based on the measure of intersection the frequency response makes with the geometric region. It functions to drive the state from any initial condition to the zero state, which indicates that the frequency response does not intersect any forbidden FD set.

This paper suggests a hybrid, frequency domain-embedded, state space approach that is capable of designing compensators of arbitrary order for a large class of systems. The design procedure is applicable to a wide class of systems provided that the stability/performance conditions are geometrically-interpreted in FD. The design approach is developed and a proof of its ability to converge, if possible, to the tuning parameter set that satisfy the desired conditions is provided. A set of design examples are supplied to demonstrate the applicability of the approach to a large class of systems.

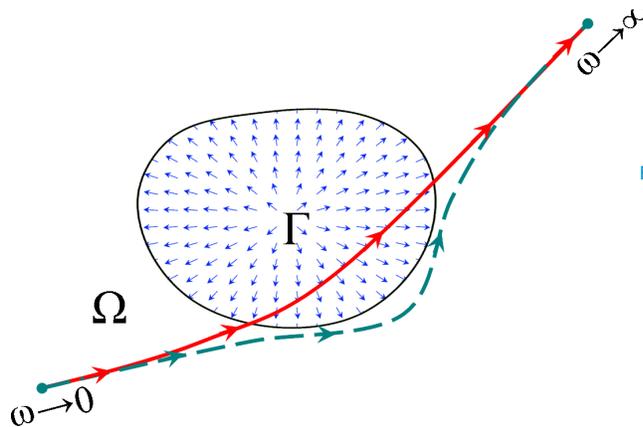

FIGURE-2: Repelling forces acting on the frequency response of a system.

## II. PROBLEM STATEMENT

This work assumes that the behavior of a system can be adjusted through L tuning parameters ($\Lambda \in R^L$, $\Lambda = [\gamma_1 \ \gamma_2 \ .. \ \gamma_L]^T$). These parameters can either be naturally present in the system or artificially introduced by a compensation unit used to produce a control input in aim with a given stability/performance criterion. The design method assumes the presence of a frequency domain assessment function (FDAF), $Q(\omega, \Lambda)$. The state of this function is used to judge system compliance with the performance specifications. The method also requires the availability of an admissible region in the complex plane $\Psi \subset C$, which encodes the desired system specifications. If $Q(\omega, \Lambda) \in \Psi \quad \forall \omega$, then the system satisfies the specifications, else, the system fail to satisfy the specifications.

The control design process is defined as finding a value for the tuning parameters that would confine the FDAF to the admissible zone in the complex plane. This is carried-out by considering $\Lambda$ as a state vector and designing the 1$^{st}$ order nonlinear dynamical system

$$\dot{\Lambda} = \mathbf{H}(\Lambda, Q(\omega, \Lambda), \Psi) \qquad \Lambda(0) \in R^L \qquad (1)$$

So that $\lim_{t \to \infty} \Lambda(t) \in \Lambda_\Psi$

where $\Lambda_\Psi = \{\Lambda : Q(\omega, \Lambda) \in \Psi\}$

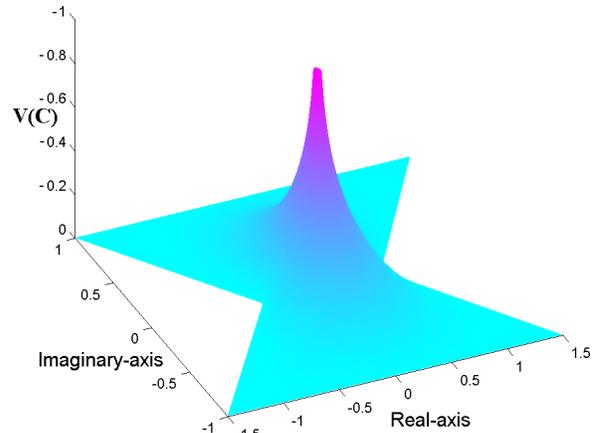

FIGURE-3: The distance function V

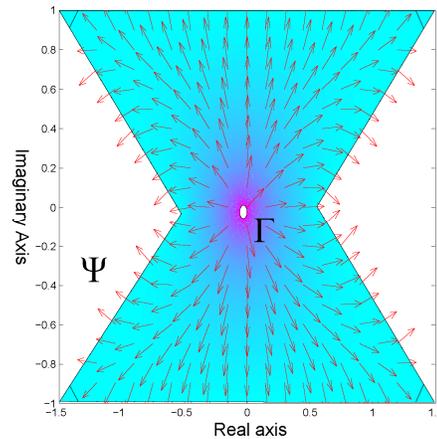

FIGURE-4: Repelling force corresponding to V in figure-3

## III. THE SUGGESTED METHOD

The virtual action **F** that is covering the region $\Gamma$ in **C** ($\Gamma = \mathbf{C} - \Psi$) is the gradient of an underlying scalar function V ($\mathbf{F} = -\nabla V$). V measures the extent a point violates the boundary of $\Gamma$. **F** maps each point in **C** to a force ($\mathbf{F}(P) : \mathbf{C} \to R^2$, $\mathbf{F}(P) = [F_r(P) \ F_I(P)]^T$, ) in such a way that creates a repelling action in $\Gamma$ and zero action in $\Psi$. If at a certain frequency a point of $Q(\omega, \Lambda)$ is present in $\Gamma$ ($Q(\omega, \Lambda) \cap \Gamma \neq \varphi \quad \forall \omega$), i.e. the system is violating the stability/performance constraints, this virtual force acts to drive that point in $\Gamma$ to a point in $\Psi$ (figure-2) so that for the dynamical system

$$\begin{bmatrix} \dot{P}_r \\ \dot{P}_I \end{bmatrix} = \begin{bmatrix} F_r(P) \\ F_I(P) \end{bmatrix} \qquad P(0) \in \Gamma \qquad (2)$$

$$\lim_{t \to \infty} P(t) \in \Psi$$

where $P_r$=re(P) and $P_I$=Im(P).

Although existing frequency domain stability criteria use stability/performance regions with simple geometry (e.g. circles or straight lines), V and **F** can be synthesized in a provably-correct manner [28] that satisfies (2) for any geometry of the stability/performance regions (figures-3,4) using the boundary value problem (3)

$$\nabla^2 V(\mathbf{C}) = 0 \qquad \forall \mathbf{C} \qquad (3)$$

subject to $V(\mathbf{C}) = 0$ *at* $\mathbf{C} = \partial\Gamma$ and $V(q) = -1$, $q \in \Gamma$.

Since **F** can affect one point only, the state of the continuous FDAF spectrum ($Q(\omega, \Lambda)$) is represented by the collective state of its points at frequency samples ($\omega_i$ i=1:N). Since all natural systems are finite energy, which in turn makes them band-limited, the samples may be taken in accordance with the sampling theorem for perfect reconstruction of $Q(\omega, \Lambda)$ from its samples. However, that is not necessary since one may concentrate only on the regions in **C** where $\Gamma$ is present in order to steer the FDAF away from them. The vector $\mathbf{F}_\Omega(\Omega)$ is an extended virtual repelling action at every sampling frequency in the complex plane

$$\mathbf{F}_\Omega(\Omega) = \begin{bmatrix} \mathbf{F}(Q(\omega_1, \Lambda)) \\ \mathbf{F}(Q(\omega_2, \Lambda)) \\ . \\ . \\ \mathbf{F}(Q(\omega_N, \Lambda)) \end{bmatrix} \qquad (4)$$

where $\Omega = \begin{bmatrix} \omega_1 & \omega_2 & . & \omega_N \end{bmatrix}^T$ is the sampling frequency vector. $\mathbf{F}_\Omega(\Omega)$ is in essence a state vector that indicates whether or not the system is complying with the stability/performance constraints. If the value of this vector is identically zero, the system is in a state of compliance with the specifications, else it is not

$$\begin{bmatrix} |F_\Omega| > 0 & \text{system violates constraints} \\ |F_\Omega| = 0 & \text{system complies with constraints} \end{bmatrix} \qquad (5)$$

The individual components of $\mathbf{F}_\Omega(\Omega)$ may be used as the coordinates of a 2N-dimensional state space ($\mathcal{F}$).

Since, by design, the force $\mathbf{F}_\Omega(\Omega)$ repels the FDAF from the $\Gamma$ region, its value will decay to zero. This means that for the state space system, $\mathbf{F}_\Omega(\Omega)$ is both a state variable and a state evolution action (6):

$$\dot{\mathbf{F}}_\Omega = \mathbf{F}_\Omega \qquad (6)$$

$$\lim_{t \to \infty} \mathbf{F}_\Omega = 0.$$

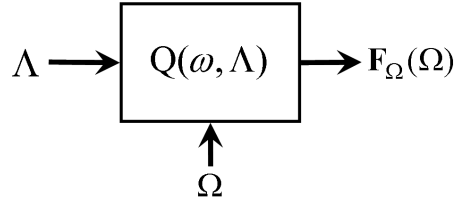

FIGURE-5: The generation of a tuning parameter-dependant state vector.

The usability of $\mathbf{F}_\Omega$ is limited since a change in the state of $Q(\omega, \Lambda)$ can only be induced through changes in the state of the parameter vector $\Lambda$ of the compensator (figure-5). The desired action $\mathbf{F}_\Omega(\Omega)$ the compensator is expected to exert on $Q(\omega, \Lambda)$ needs to be converted into an equivalent action that $\Lambda$ exerts on the FDAF. To achieve this, an L-dimensional state space whose coordinates are the individual components of the parameter space vector $\Lambda$ ($\mathcal{L}$) has to be considered. Under the influence of the transformation in (4), the parameter space forms an L-dimensional image subspace ($\mathcal{F}_\Lambda$) in $\mathcal{F}$ (figure-6). **I**f a solution exists yielding a value for $\Lambda$ that is able to place $Q(\omega, \Lambda)$ in $\Psi$, or equivalently drive $\mathbf{F}_\Omega$ to zero, $\mathcal{F}_\Lambda$ will be connected and pass through the origin of $\mathcal{F}$. The admissible action that the compensator can use to influence the frequency response of the system is the projection of $\mathbf{F}_\Omega$ in the $\mathcal{F}_\Lambda$ sub-space. This action is used to drive the evolution of $\Lambda$ in the $\mathcal{L}$ space.

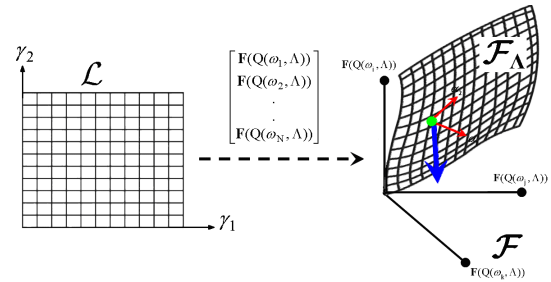

FIGURE-6: Mapping of parameter space into the frequency sample space.

Computing the projection $\mathbf{F}_\Omega$ makes on $\mathcal{F}_\Lambda$ requires the construction of a complete orthogonal set of vectors that are tangent to $\mathcal{F}_\Lambda$. It is possible to build these vectors from the columns of the extended Jacobian matrix at each sampling frequency ($\mathbf{J}_\Omega(\Omega)$)

$$\mathbf{J}_\Omega(\Omega) = \begin{bmatrix} \mathbf{J}(\omega_1) \\ \mathbf{J}(\omega_2) \\ . \\ . \\ \mathbf{J}(\omega_N) \end{bmatrix} \quad (7)$$

where

$$\mathbf{J}(\omega) = \begin{bmatrix} \dfrac{\partial \mathrm{Re}(Q(\omega,\Lambda))}{\partial \Lambda} \\ \dfrac{\partial \mathrm{Im}(Q(\omega,\Lambda))}{\partial \Lambda} \end{bmatrix} \quad (8)$$

The projection of $\mathbf{F}_\Omega$ on $\mathcal{F}_\Lambda$ ($\mathbf{F}_\Lambda$)

$$\mathbf{F}_\Lambda = \mathbf{J}_\Omega^T(\Omega)\mathbf{F}_\Omega(\Omega) \quad (9)$$

maybe used to drive the evolution of the parameter vector $\Lambda$. The control design method is expressed as the nonlinear state space system:

$$\dot{\Lambda} = \mathbf{J}_\Omega^T(Q(\Omega,\Lambda))\mathbf{F}_\Omega(Q(\Omega,\Lambda)) \qquad \Lambda(0) = \Lambda_0 \quad (10)$$

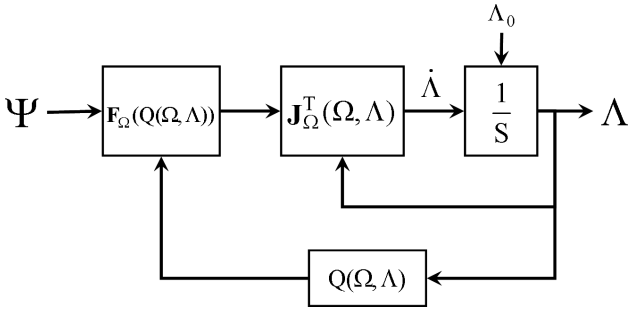

FIGURE-7: Block diagram of the control synthesis method.

Since the system in (6) is able to drive $\mathbf{F}_\Omega$ to zero and $Q(\omega,\Lambda)$ to $\Psi$, the projection of $\mathbf{F}_\Omega$ on $\mathcal{F}_\Lambda$ should be able to drive $\Lambda$ to a value that steers $\mathbf{F}_\Omega$ to zero. The block diagram describing the design process is shown in figure-7.

The following proposition demonstrates the ability of system (10) to converge globally to a set of tuning parameters that make the system comply with the desired constraints.

Proposition-1: Consider the state space system:

$$\dot{\Lambda} = \mathbf{J}_\Omega^T(Q(\Omega,\Lambda))\mathbf{F}_\Omega(Q(\Omega,\Lambda)) \qquad \Lambda(0) = \Lambda_0$$

If $\mathbf{J}_\Omega$ is full column rank for any nonzero $\mathbf{F}_\Omega$ and if for $Q(\omega,\Lambda) \cap \Gamma \neq \phi$ there exist a $Q(\omega_i,\Lambda) \cap \Gamma \neq \phi$, then

$$\lim_{t\to\infty} \mathbf{F}_\Omega(\Lambda(t)) = 0, \quad (11)$$

and

$$\lim_{t\to\infty} \Lambda(t) \in \Lambda_\Psi$$

where $\Lambda_\Psi$ is the subset of $\Lambda$ ($\Lambda_\Psi \subset \Lambda$) so that $Q(\omega,\Lambda_\Psi) \in \Psi \quad \forall \omega$ ($\Lambda_\Psi = \{\Lambda : Q(\omega,\Lambda) \in \Psi\}$).

Proof: Consider the positive scalar function V(P) that is constructed so that

$$\begin{aligned} V(P) &= 0 \quad \forall P \in \Psi \\ and \quad \nabla V(P) &= -\mathbf{F}(P) \quad \forall P \in \Gamma \end{aligned} \quad (12)$$

where $\mathbf{F}$ satisfies (2). As a result

$$\begin{aligned} V(P) &> 0 \quad \forall P \in \Gamma \\ and \quad \nabla V(P) &= 0 \quad \forall P \in \Psi \end{aligned} \quad (13)$$

Consider the extended function $V_\Omega(\Lambda)$

$$V_\Omega(\Lambda) = \sum_{i=1}^{N} V(Q(\omega_i,\Lambda)) \quad (14)$$

Then

$$\begin{bmatrix} V_\Omega(\Lambda) = 0 & if & \Lambda \in \Lambda_\Psi \\ V_\Omega(\Lambda) > 0 & if & \Lambda \notin \Lambda_\Psi \end{bmatrix} \quad (15)$$

Using the chain rule to compute the derivative of $V_\Omega(\Lambda)$ with respect to time we have:

$$\begin{aligned} \dfrac{d V_\Omega(\Lambda)}{dt} &= \sum_{i=1}^{N} \nabla V^T \cdot \dfrac{dQ(\omega_i,\Lambda)}{d\Lambda} \cdot \dfrac{d\Lambda}{dt} \\ &= -\sum_{i=1}^{N} F^T(Q(\omega_i,\Lambda)) \cdot \mathbf{J}(Q(\omega_i,\Lambda)) \dfrac{d\Lambda}{dt} \end{aligned} \quad (16)$$

The above expression may be written as

$$\dfrac{d V_\Omega(\Lambda)}{dt} = -\mathbf{F}_\Omega^T \cdot \mathbf{J}_\Omega \cdot \dfrac{d\Lambda}{dt} \quad (17)$$

If the derivative of $\Lambda$ is selected as in (10)

$$\dfrac{d V_\Omega(\Lambda)}{dt} = -\mathbf{F}_\Omega^T \cdot \mathbf{J}_\Omega \cdot \mathbf{J}_\Omega^T \mathbf{F}_\Omega = -\left(\dfrac{d\Lambda}{dt}\right)^T \dfrac{d\Lambda}{dt} \quad (18)$$

The matrix $\mathbf{J}_\Omega \cdot \mathbf{J}_\Omega^T$ is always positive semi-definite. If $\mathbf{J}_\Omega$ is full column rank, $d\Lambda/dt$ is non-zero if $\mathbf{F}_\Omega$ is non-zero, the derivative of $V_\Omega(\Lambda)$ is strictly decreasing and

$$\lim_{t\to\infty} V_\Omega(\Lambda(t)) = 0. \quad (19)$$

This implies that $\lim_{t \to \infty} \mathbf{F}_\Omega(\Lambda(t)) = 0$,

and $\lim_{t \to \infty} \Lambda(t) \in \Lambda_\Psi$.

In other words, the system converges to the set of tuning parameter that makes the system comply with the requirements.

### A. A note on the design method:

The dynamics of the design system controlling the evolution of the tuning parameters is expressed as (20)

$$\dot{\Lambda} = \begin{bmatrix} \dot{\gamma}_1 \\ \dot{\gamma}_2 \\ \vdots \\ \dot{\gamma}_L \end{bmatrix} = \begin{bmatrix} g(\omega_1) \cdot \frac{\partial g_c(\omega_1,\Lambda)}{\partial \gamma_1} & g(\omega_2) \cdot \frac{\partial g_c(\omega_2,\Lambda)}{\partial \gamma_1} & \cdots & g(\omega_N) \cdot \frac{\partial g_c(\omega_N,\Lambda)}{\partial \gamma_1} \\ g(\omega_1) \cdot \frac{\partial g_c(\omega_1,\Lambda)}{\partial \gamma_2} & g(\omega_2) \cdot \frac{\partial g_c(\omega_2,\Lambda)}{\partial \gamma_2} & \cdots & g(\omega_N) \cdot \frac{\partial g_c(\omega_N,\Lambda)}{\partial \gamma_2} \\ \vdots & \vdots & & \vdots \\ g(\omega_1) \cdot \frac{\partial g_c(\omega_1,\Lambda)}{\partial \gamma_L} & g(\omega_2) \cdot \frac{\partial g_c(\omega_2,\Lambda)}{\partial \gamma_L} & \cdots & g(\omega_N) \cdot \frac{\partial g_c(\omega_N,\Lambda)}{\partial \gamma_L} \end{bmatrix} \begin{bmatrix} f_{\omega_1} \\ f_{\omega_2} \\ \vdots \\ f_{\omega_N} \end{bmatrix} \quad (20)$$

$$= \sum_{i=1}^{N} f_{\omega_i} \cdot g(\omega_i) \cdot \begin{bmatrix} g(\omega_i) \cdot \frac{\partial g_c(\omega_i,\Lambda)}{\partial \gamma_1} \\ g(\omega_i) \cdot \frac{\partial g_c(\omega_i,\Lambda)}{\partial \gamma_2} \\ \vdots \\ g(\omega_i) \cdot \frac{\partial g_c(\omega_i,\Lambda)}{\partial \gamma_L} \end{bmatrix}$$

where $f_{\omega_i}$ is the i'th component of $\mathbf{F}_\Omega$, $g(\omega_i)$ is the i'th component of the vector describing the frequency spectrum of the uncompensated system whose response has to be shaped and $g_c(\omega_i,\Lambda)$ is the i'th component of the vector describing the frequency response of the compensator system. Notice that the design method has a connectionist, parallel-distributed nature. Each virtual force at each frequency sample has the potential of modifying all the tuning parameters (figure-8). Notice that the number of virtual force frequency samples (N) compared to the number of tuning parameters used (L) is quite high (N>>L). Since each force can affect all tuning parameters, it is with probability one that the design method will work if the design problem is well-posed.

The convergence condition of $\mathbf{J}_\Omega$ in (7) being full column rank is by no means stringent. However, there are situations when convergence fails due to the order of the compensator being lower than what is needed to shape the system's frequency response. Due to the fact that the matrix $\mathbf{J}_\Omega \mathbf{J}_\Omega^T$ is always positive semi-definite, divergence is ruled-out and design failure manifest itself as zero change ($\dot{\Lambda} = 0$) in the value of the tuning parameters (21). When this happens the individual vector actions induced in parameter space by the individual components of the virtual force ($\Delta_\Lambda(\omega_i)$) are in an equilibrium position (i.e. linearly dependant) and cancel-out (figure-9).

The convergence problem can be easily solved by increasing the dimensionality of $\mathcal{L}$. This implies increasing both the order of the compensating system and the number of tuning parameters it has. Note that the number of actions in $\mathcal{L}$ ($\Delta_\Lambda(\omega_i)$'s) is determined by the number of sampling frequencies. The scaling factors of the actions are only determined by the value of the sampling frequencies. Therefore an increase in dimensionality of $\mathcal{L}$ reduces the likelihood of $\Delta_\Lambda(\omega_i)$'s cancelling each other by falling in an equilibrium configuration. Consequently, halt in the change in the tuning parameters can only happen when all the components of $\mathbf{F}_\Omega$ are zero and the design process met the desired criterion.

$$\dot{\Lambda} = \begin{bmatrix} 0 \\ 0 \\ \vdots \\ 0 \end{bmatrix} = \sum_{i=1}^{N} f_{\omega_i} \cdot g(\omega_i) \cdot \begin{bmatrix} g(\omega_i) \cdot \frac{\partial g_c(\omega_i,\Lambda)}{\partial \gamma_1} \\ g(\omega_i) \cdot \frac{\partial g_c(\omega_i,\Lambda)}{\partial \gamma_2} \\ \vdots \\ g(\omega_i) \cdot \frac{\partial g_c(\omega_i,\Lambda)}{\partial \gamma_L} \end{bmatrix} = \sum_{i=1}^{N} \Delta_\Lambda(\omega_i) \quad (21)$$

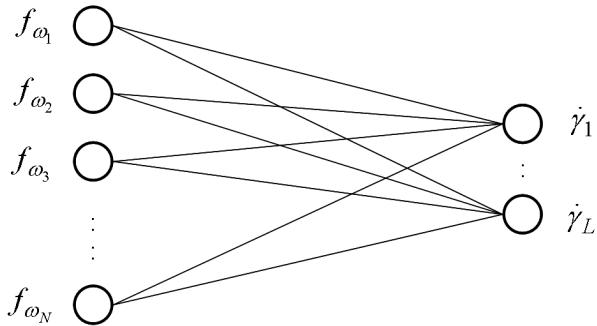

FIGURE-8 The connectionist, parallel-distributed nature of the design method

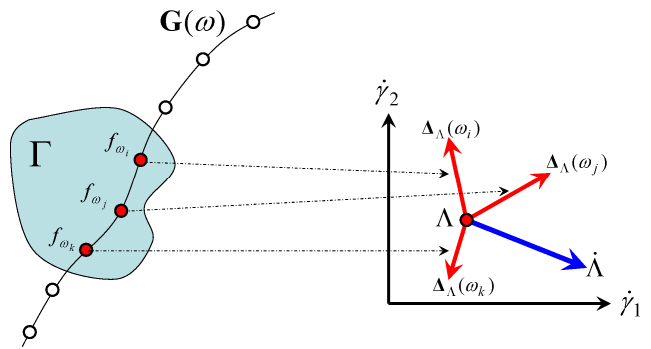

FIGURE-9: Individual vector actions induced in parameter ($\mathcal{L}$) space by $\mathbf{F}_\Omega$

## IV. SIMULATION RESULTS

The following series of examples demonstrate the ability of the suggested method to design controllers for linear and nonlinear systems alike. The system nature does not matter provided that the stability or performance criterion of the system has a geometric interpretation in the frequency domain. The method can place constraints on the parameters of the compensators. This maybe used, among other things, to restrict the zeros of the compensators to be non-negative to prevent the system becoming non-minimum phase or to prevent pole cancellation with the original system.

### A. Third order system:

In this example the design method is tested on a third order system with transfer function

$$G(S) = \frac{1}{S^3} \qquad (22)$$

The system is cascaded with a first order compensator with Z and P as tuning parameters. Its transfer function is

$$G_c(S,P,Z) = \frac{S+Z}{S+P} \qquad (23)$$

The cascade is placed in a unity feedback configuration. The method is required to compute the tuning parameters so that the frequency response of the feed-forward transfer function $(G(S)G_c(S,P,Z))$ does not enclose -1 on the real axis. It is also required to stay outside a circle with a radius 0.75 surrounding -1. The initial values used by the design procedure are: Z=0.45, P=0.55. The procedure converges to Z=0 and P=1.67. The evolution curve of the tuning parameters, the open loop (OL) frequency response and the closed loop (CL) step response are shown in figures-10, 11 and 12 respectively. As expected, it was not possible to stabilize the system since (22) is impossible to stabilize using a 1st order compensator.

A second order compensator is used instead of the first order one. Its transfer function is:

$$G_c(S,P1,Z1,P2,Z2) = \frac{S+Z1}{S+P1} \cdot \frac{S+Z2}{S+P2} \qquad (24)$$

The initial conditions used are: Z1=0.45, P1=0.55, Z2=0.40, P2=0.60. They produce an unstable closed loop step response (figure-13). The design algorithm is required to stabilize the system and keep the OL frequency response outside a circle of radius 0.75 surrounding -1 on the real axis.

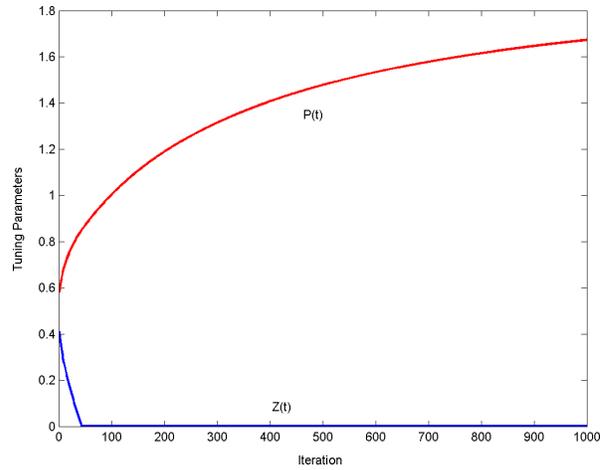

FIGURE-10  Parameter evolution, 3rd order system with 1st order compensator.

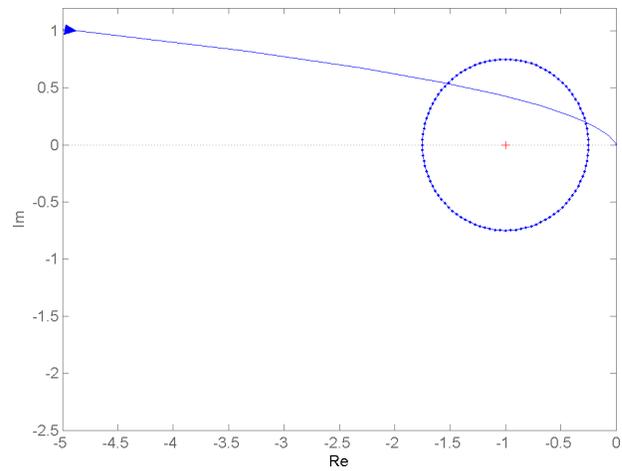

FIGURE-11  OL frequency response, 3rd order system with 1st order compensator.

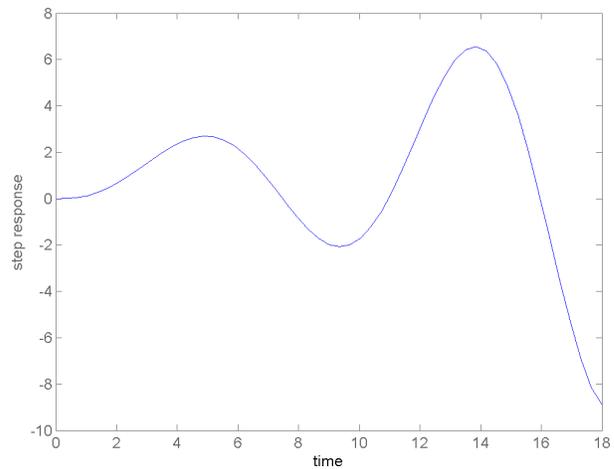

FIGURE-12: CL step response respectively, 3rd order system with 1st order compensator.

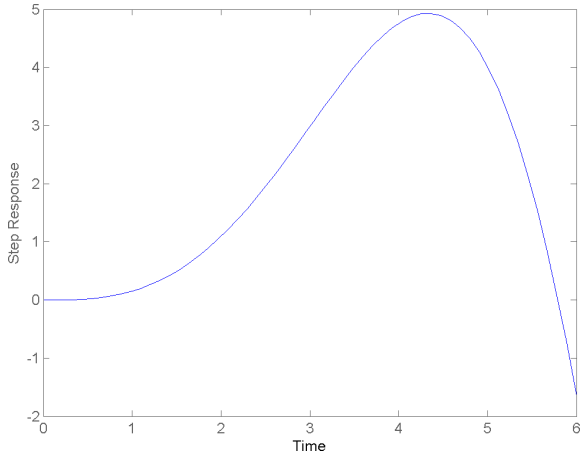

FIGURE-13  unstable step response of 3$^{rd}$ order system with 2$^{nd}$ order compensator, initial choice of parameters.

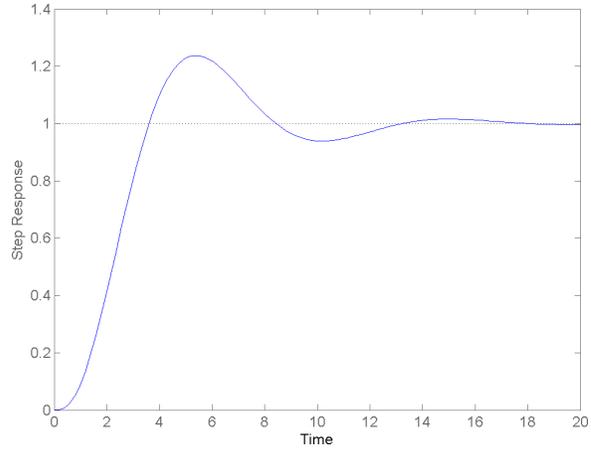

FIGURE-16  CL step response respectively, 3$^{rd}$ order system with 2$^{nd}$ order compensator.

The method converged to Z1=0.0, P1=1.27, Z2=0.0, P2=1.29. The evolution of the tuning parameters, the OL frequency response and the CL step response are shown in figures: 14, 15 and 16 respectively. As can be seen, the system successfully met the design requirements.

The above example is repeated with constraints on the zeros of the compensator not be less than 0.05 to prevent pole zero cancellation with original system poles. Figure-17 shows the evolution of the compensator's parameters and Figure-18 shows the step response. The design converged to Z1=Z2=.05 and P1=1.348, P2=1.365. As can be seen the restriction on parameters had little effect on the design.

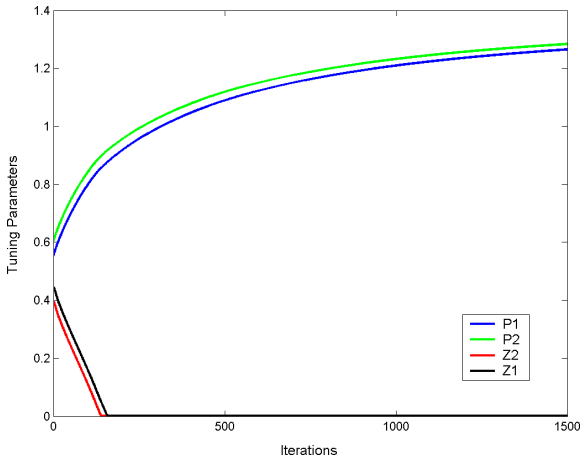

FIGURE-14  Parameter evolution, 3$^{rd}$ order system with 2$^{nd}$ order compensator.

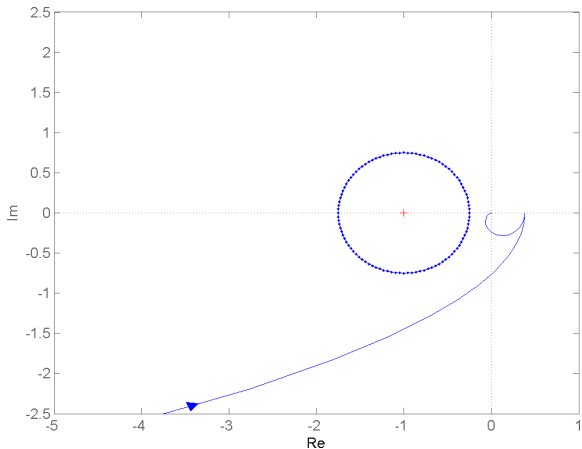

FIGURE-15  OL frequency response, 3$^{rd}$ order system with 2$^{nd}$ order compensator.

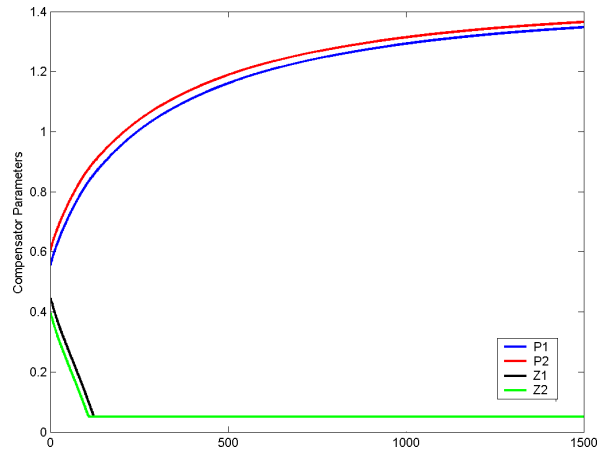

FIGURE-17  Design in figure-16 repeated with zeros of compensator restricted not to go below 0.05.

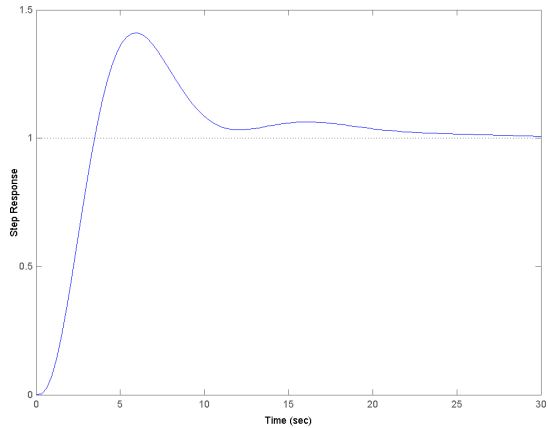

FIGURE-18  Design in figure-16 repeated with zeros of compensator restricted not to go below 0.05.

Figure-19 shows the step responses for different phase margins corresponding to circles of radii 0.2, 0.4, 0.6 and 0.8. As expected, the bigger the phase margin, the lower is the overshoot and settling time.

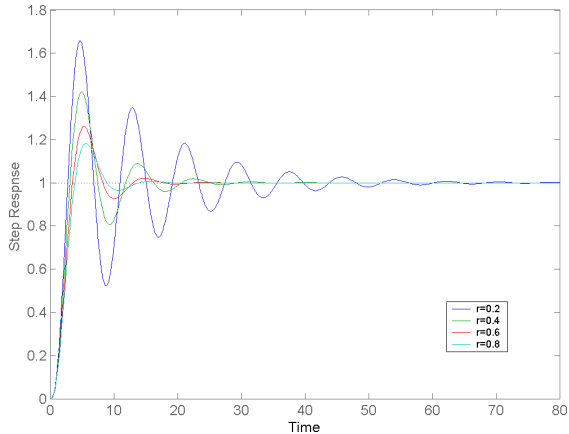

FIGURE-19  The more the radius of the circle surrounding (-1,0) is increased, the faster is the response and the lower is the overshoot.

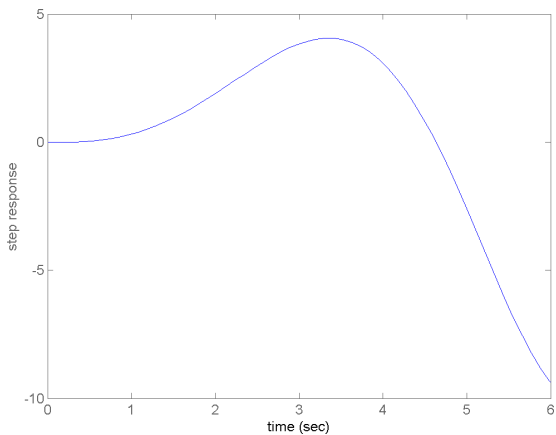

FIGURE-20  unstable CL step response of the system in (25) with 1$^{st}$ order compensator, initial choice of parameters.

### B. Second order system with delay:

In this example the design method is tested on a second order system with delay whose transfer function is

$$G(S) = \frac{1}{S^2} \cdot e^{-T \cdot S} \qquad (25)$$

The system is cascaded with a first order compensator with Z and P as tuning parameters. Its transfer function is

$$G_c(S) = \frac{S+Z}{S+P} \qquad (26)$$

The design procedure is required to stabilize the system when the delay is T=0.1 seconds. It is also required to keep the OL frequency response outside a circle of radius 0.7 surround -1 on the real axis. The initial values for the design program is initialized with Z=1.0 and P=0.1 which produces an unstable CL step response (figure-20)

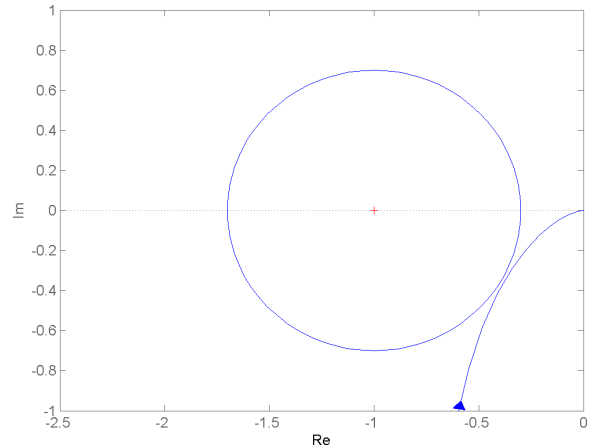

FIGURE-21  OL frequency response of the system in (25), delay=0.1 second

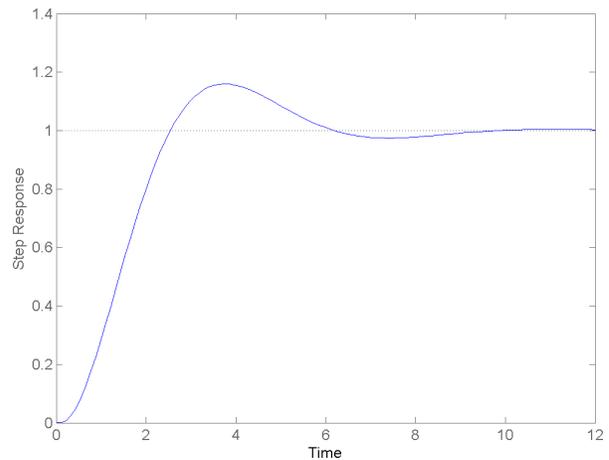

FIGURE-22  CL step response respectively of the system in (25), delay=0.1 second

VOLUME. 00 2021                                                                                                                                                              9

The design system converged to P=1.01 and Z= 0. The OL frequency response and the CL step response are shown in figures-21 and 22 respectively.

The time delay was varied (T=0, 0.1, 0.2 and 0.3 seconds) while keeping the above specifications the same. The OL frequency response and CL step response corresponding to these delays are shown respectively in figure-23 and 24. The radius of the circle around -1 for which no intersection with the frequency response happens started shrinking after a delay of 0.3 sec. until it was not possible to stabilize the system for delays 0.6 sec. and above.

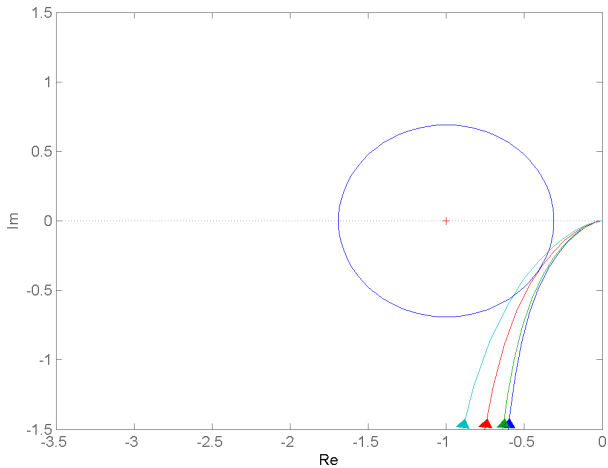

FIGURE-23 OL frequency response of the system in (25) for different delays. The higher the delay, the more is the overshoot and settling time.

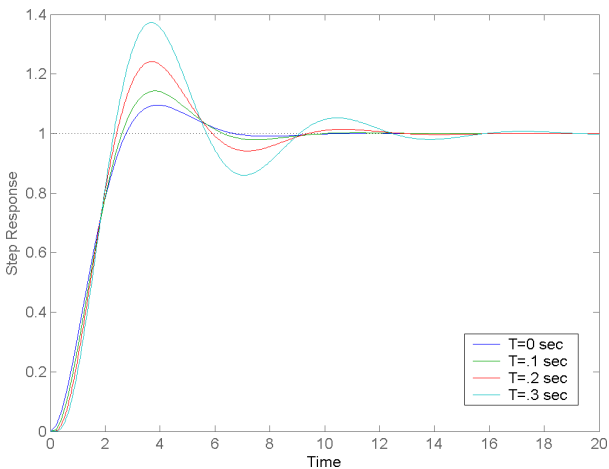

FIGURE-24 CL step response respectively of the system in (25) for different delays. The higher the delay, the more is the overshoot and settling time.

### C. 2nd order system with delay & sector nonlinearity:

A practical type of systems consists of a linear system cascaded with a static nonlinearity (figure-25). This nonlinearity is usually a first and third sector nonlinearity (figure-26) that is bounded between two lines with positive slopes K1 and K2. Such nonlinearities include saturation, deadzone and variable gain among others. Stability for such systems may be guaranteed using the circle criterion [20,21,22]. The criterion requires that the frequency response of the linear part of the system (assuming it has no poles or zeros in the right half of the S-plane) to stay outside the circle (figure-26).

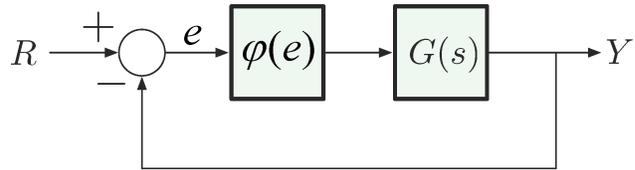

FIGURE-25 A system with OL consisting of a linear system cascaded with a nonlinearity

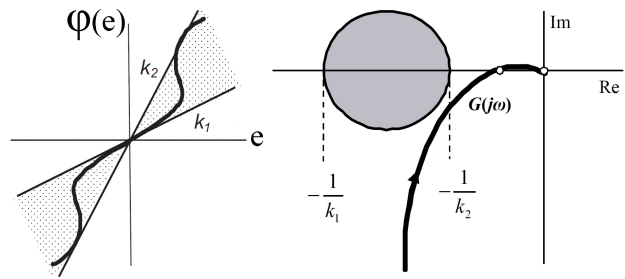

FIGURE-26 Sector nonlinearity and the circle stability criterion.

The linear system used (G=Gs·Gc) is a cascade of a second order linear system (Gs) with delay (T=0.2 second)

$$G_s(S) = \frac{1}{S^2} \cdot e^{-T \cdot S} \quad (27)$$

and a second order compensator (Gc) with two poles and zeros as the tuning parameters

$$G_c(S, P1, Z1, P2, Z2) = \frac{S+Z1}{S+P1} \cdot \frac{S+Z2}{S+P2} \quad (28)$$

The nonlinearity used is a sinusoidally varying (29) gain that is sandwiched between two lines with slopes 1 and 0.5

$$\varphi(x) = \frac{3 \cdot x + x \cdot \sin(5 \cdot x)}{4}. \quad (29)$$

The initial values selected for the tuning parameters are P1=.1, Z1=1, P2=.2, Z2=.9 which produces an unstable response. The virtual repelling forces are designed to push G(ω) away from the stability circle as much as possible. The final value of the parameters is: Z1 = 0.2661, P1 = 0.6858, Z2 = 0, P2 = 0.7368 and their evolution curve is shown in figure-27.

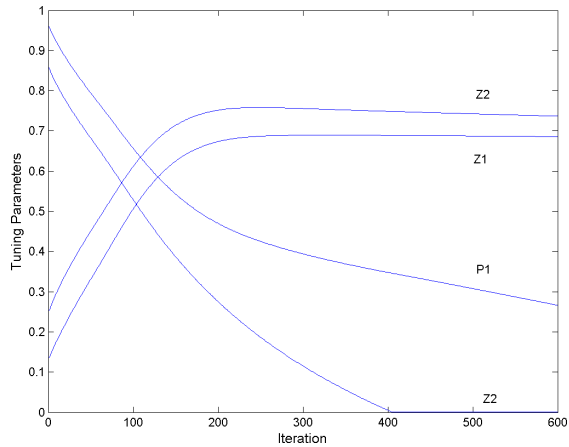

FIGURE-27  Parameters for the system in (27) with compensator in (28).

The design procedure kept the frequency response of G(S) significantly away from the stability circle. The OL frequency response and CL step response of the system are shown respectively in figure-28 and 29.

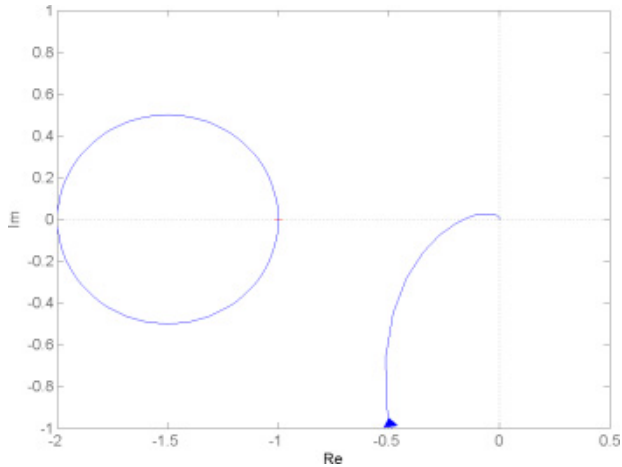

FIGURE-28  The OL frequency response of the system in 27-29.

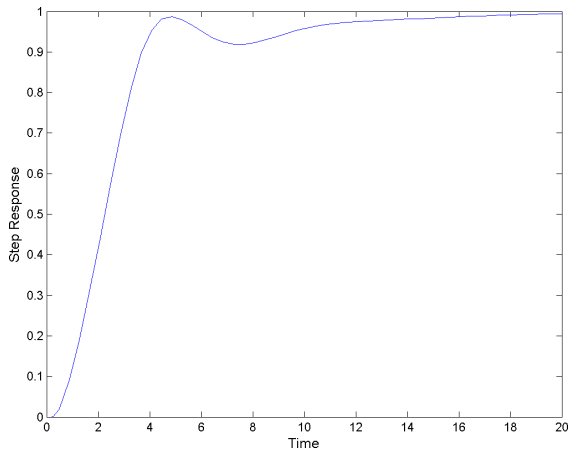

FIGURE-29  CL step response of the system in 27-29.

$$\varphi(x) = \begin{bmatrix} -0.2 & x < -0.2 \\ x & -0.2 \leq x \leq +0.2 \\ +0.2 & x > 0.2 \end{bmatrix} \quad (30)$$

The design based on the Popov frequency domain stability criterion is not sensitive to the specific shape of the nonlinearity. It is only sensitive to linear bounds on this nonlinearity. The same design for the sinusoidal nonlinearity can also guarantee stability for the saturation nonlinearity (30) that limits the input between (-0.2, 0.2). The step response with saturation nonlinearity instead of the sinusoidal nonlinearity is shown below in figure-30.

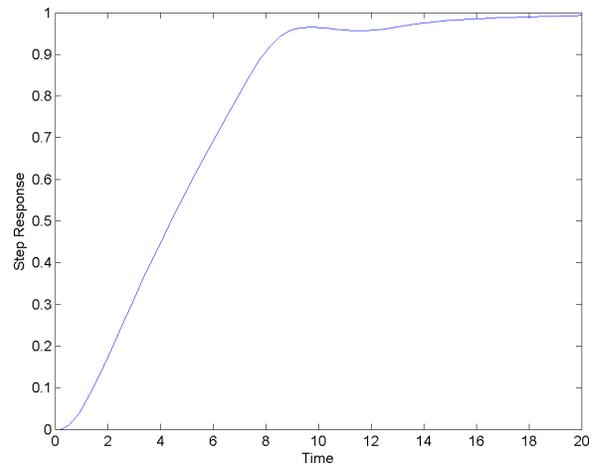

FIGURE-30  The CL step response of the system in figure-27 with saturation used as a nonlinearity.

### D. Multi-input, Multi-output (MIMO) systems:

Stabilizing MIMO systems is a challenging and active area of research even when only linear systems are concerned. The following demonstrates clearly the strong potential of the suggested approach not to only handle linear MIMO systems with practical artifacts such as delays or saturation, but to stabilize the system also in a decentralized manner [26,27].

As mentioned before, the suggested design approach is invariant to the system's nature if the stability/performance criterion has a geometric interpretation in frequency domain. The suggested method is applicable to MIMO systems since the circle stability criterion was extended to this case [23,24,25]. In the followings a procedure that capitulate on the properties of the suggested approach is provided that would convert the design of a stabilizing controller for a MIMO system that contains M plants with M individual inputs into the design of M controllers for single input single output systems.

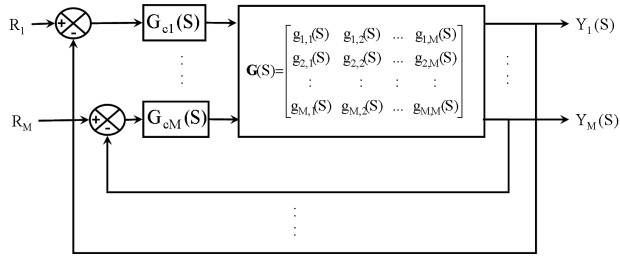

FIGURE-31 Decentralized control of a MIMO system.

Consider a MIMO system with M inputs and M outputs (figure-31). It is required to design M compensators ($G_{ci}(S,\Lambda_i)$) that can stabilize the system. The compensators generate the i'th control signal by acting only on the error corresponding to the i'the output (31)

$$U_i(S) = G_{ci}(S,\Lambda_i) \cdot (R_i - Y_i(S)) \quad (31)$$

The equivalent forward transfer function (EFTF, $Q_i(S)$) connecting an error channel $E_i(S)$ to the corresponding output $Y_i(S)$ (32) is

$$Q_i(S) = \left[ g_{i,i}(S) - \sum_{\substack{j=1 \\ j \neq i}}^{M} G_{cj}(S,\Lambda_j) \cdot g_{i,j}(S) \cdot g_{j,i}(S) \right] \cdot G_{c_i}(S,\Lambda_i) \quad i=1,..M \quad (32)$$

Therefore, if $G_{ci}(S,\Lambda_i)$ (i=1,..,M) can be found so that the individual compensated frequency response of each $Q_i(S)$ is outside the unit circle whose center is -1 on the real axis (assuming stable and minimum phase transfer function), the MIMO system can be stabilized in a decentralized manner.

The method can easily accomplish the above by consolidating the repelling virtual forces generated by the compensated frequency responses of all the elements into one virtual force. This force is used to steer the evolution of the tuning parameters of the i'th compensator.

The following example illustrates the capabilities of the method. Consider the two input two-output transfer function in (33)

$$G(s) = \begin{bmatrix} \dfrac{1}{S^3} & \dfrac{1}{S^2} \\ \dfrac{1}{S} & \dfrac{e^{-0.2 \cdot S}}{S^2} \end{bmatrix} \quad (33)$$

A second order compensator (two poles and two zeros) is used at each input to stabilize the system. The compensators computed by the method for the first and second error channels are:

$$G_{c1}(S) = \frac{S+.05}{S+1.272} \cdot \frac{S+.05}{S+1.302},$$

$$G_{c2}(S) = \frac{S+.05}{S+1.34} \cdot \frac{S+.05}{S+1.35} \quad (34)$$

The EFTFs are

$$Q_1(S) = \left[ \frac{1}{S^3} - \frac{G_{c2}(S)}{S^3} \right] \cdot G_{c1}(S),$$

$$Q_2(S) = \left[ \frac{e^{-0.2 \cdot S}}{S^2} - \frac{G_{c1}(S)}{S^3} \right] \cdot G_{c2}(S) \quad (35)$$

The OL frequency response of the EFTF-1 and 2 are shown in figure-32.

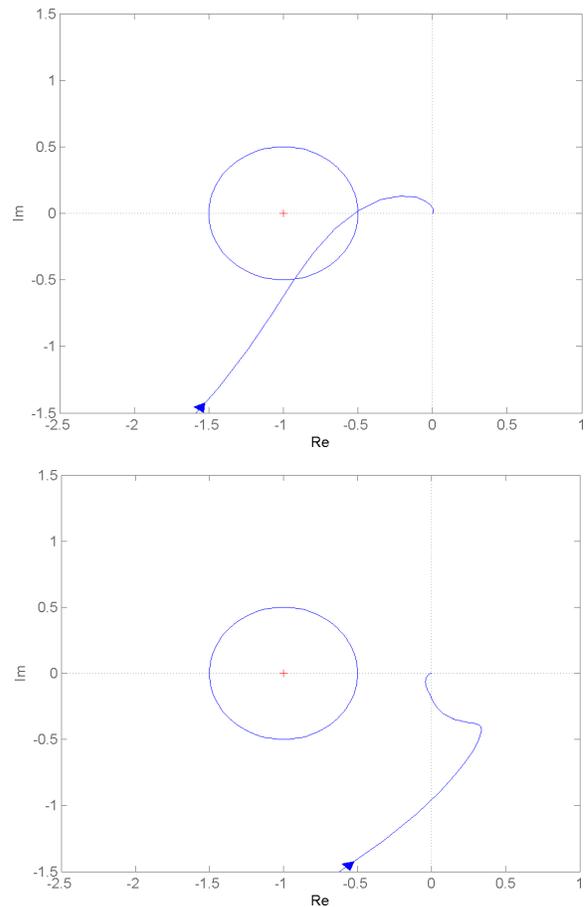

FIGURE-32 The OL frequency response of the EFTF-1 and 2 corresponding to 35.

The CL step responses of output 1 and 2 of the unity feedback MIMO system are shown in figure-33.

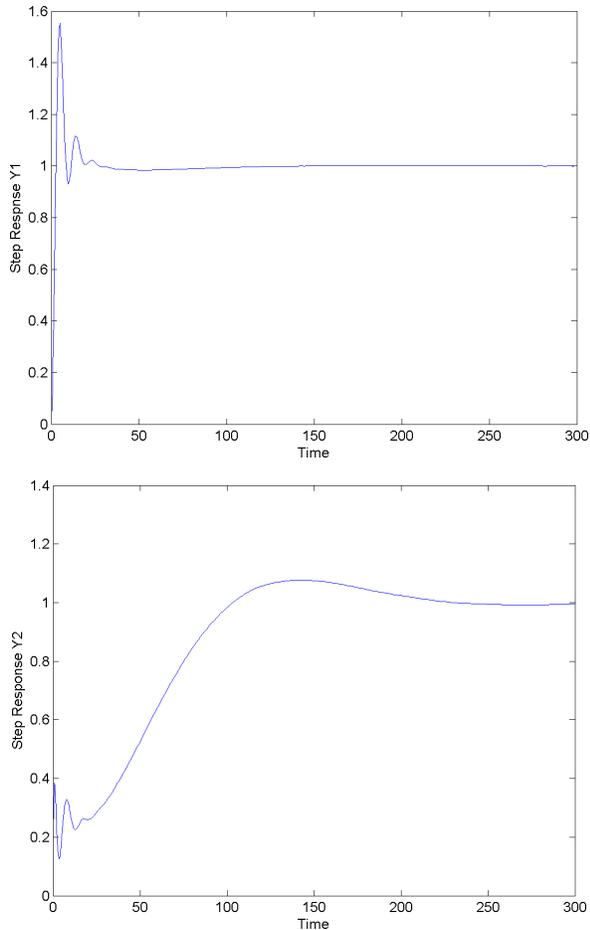

FIGURE-33 The CL step responses of output 1 and 2 of the unity feedback MIMO system corresponding to 33,34

## V. CONCLUSIONS

This paper suggests a method for converting stability and performance conditions into a controller. The design method consists of a frequency domain-embedded state space system whose states are the tuning parameters of the controller. The suggested method has the ability to treat in a unified manner linear, nonlinear, SISO and MIMO systems provided that the stability/performance criterion can be geometrically interpreted in the frequency domain. The approach has high flexibility in terms of not imposing a specific geometry on the stability/performance conditions, its ability to factor explicitly frequency in the frequency response shaping and the ability to exercise holistic influence over the frequency response through special design of the virtual force. It also does not place any restrictions on the order of the compensator used to condition the frequency response. Along with the convenience and efficiency it provides the synthesis process with, the method is expected to have a positive influence on system analysis. It may produce a feasible framework that makes it possible to fuse high-level controllers with low-level controller in a provably-correct manner.

**ACKNOWLEDGMENT**

The author would like to thank King Fahad University of Petroleum and minerals (KFUPM) for its support of this work. The author is with the center for communication systems and sensing.

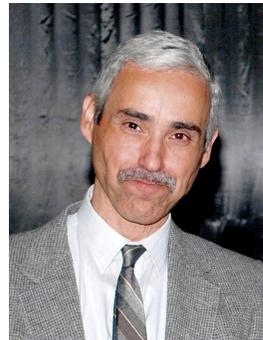

**Ahmad A. Masoud** received his B.Sc. degree in electrical engineering with a major in power systems and a minor in communication systems from Yarmouk University, Irbid, Jordan, in 1985, and his M.Sc. and Ph.D. degrees in electrical engineering both from Queen's University, Kingston, Ontario, Canada, in 1989 and 1995, respectively. He worked as a researcher with the Electrical Engineering Department, Jordan University of Science and Technology, Irbid, from 1985 to 1987. He was also a parttime assistant professor and a research fellow with the Electrical Engineering Department, Royal Military College of Canada, Kingston, from 1996 to 1998. During that time, he carried out research in digital signal processing-based demodulator design for high density, multiuser satellite systems and taught courses in robotics and control systems. He is currently an assistant professor with the Electrical Engineering Department, King Fahd University of Petroleum and Minerals, Dhahran, Saudi Arabia. His current interests include navigation and motion planning, robotics, constrained motion control, intelligent control, and DSP applications in machine vision and communication